\documentstyle[preprint,aps,eqsecnum]{revtex}
\begin{document}
\draft
\preprint{Alberta Thy 17-95 \ \ \ hep-th/9508077}
\title{Two observers calculate the trace anomaly}
\author{ A.\ Z. Capri,
M. Kobayashi\cite{address} and D.\ J. Lamb\cite{email} }
\address{ Theoretical Physics Institute, \\
Department of Physics,
University of Alberta,\\
Edmonton, Alberta T6G 2J1, Canada}
\date{\today}
\maketitle
\begin{abstract}
We adapt a calculation due to Massacand and Schmid  to the
coordinate
independent definition of time and vacuum given by Capri and Roy in
order to compute the trace anomaly for a massless scalar field in a
curved spacetime in 1+1 dimensions.  The computation which
requires only a simple regulator and
normal ordering yields the well-known result  $\frac{R}{24\pi}$  in a
straightforward manner.
\end{abstract}


\section{Introduction}
One of the more interesting results of the study of quantum field
theory
in curved spacetime is the fact that the expectation value of the
trace
of the stress tensor of a conformally coupled field does not vanish.
It
has an anomaly. This trace or conformal anomaly, as it is known, was
first
noticed by Capper and Duff \cite{3} using a dimensional
regularization scheme.
Since then many other regularization procedures have been used and
when used
correctly lead to same result \cite{4}.
Unfortunately, as anyone who has ever calculated this trace
anomaly knows, the computations required are rather lengthy and
certainly
less than illuminating.  On the other hand, if one has a particle
interpretation the problem can be handled more simply.  This fact was
first exploited by Massacand and Schmid \cite{1}.  In this paper we
adapt their
method to a computation in 1+1 dimensions using only the following
two
inputs.

\noindent 1)  The frame components of the stress tensor at a given
point are,
for two frames based at this point, related by a Lorentz
transformation.

\noindent 2)  The vacuum expectation value of the energy momentum
density
(relative to a given frame) should vanish.  Thus, the vacuum can have
pressure,
but no energy or momentum.

	In general there would remain the vexing question, ``Which
vacuum?" The answer we propose is to use the coordinate independent
definition of Capri and Roy.
	In section II we give a brief review of this construction of
the
vacuum and apply the result to a calculation of the vacuum
expectation
value  of the trace of the stress tensor in section III.  Our
conclusions are
set out in section IV.

\section{Coordinate independent definition of time and vacuum}

In a globally hyperbolic spacetime one can choose a foliation based
solely on geodesics.  Thus, given a timelike (unit) vector
$N_{\mu}(P_0)$ at
the
point $P_0$  one establishes a frame (zweibein) at $P_0$ with
components:
\begin{eqnarray}
e^{\mu\hat{0}} &=& N^\mu(P_0) \nonumber \\
e^{\mu\hat{1}} &=&  p^\mu(P_0)
\label{1}
\end{eqnarray}
where  $p^\mu(P_0)$  is a unit vector orthogonal to $N^\mu(P_0)$ at
$P_0$.  The
spacelike hypersurface (line) consiting of the geodesic through $P_0$
with
tangent vector $p^\mu(P_0)$ defines the surface  $t=0$.  The ``time"
$t$
corresponding to an arbitrary point  $P$  is the distance along a
geodesic
$P_1-P$ which intersects the line  $t=0$ orthogonally at some point
$P_1$ .
The geodesic distance $P_0-P_1$ along the line $t=0$ yields the space
coordinate $x$. These geodesic normal coordinates prove to be very
useful since
in these coordinates the metric becomes
\begin{equation}
ds^2 = dt^2 - \alpha^2(t,x) dx^2
\label{2}
\end{equation}
where
\begin{eqnarray}
\alpha(0,0) &=& 1  \nonumber \\
  \frac{\partial\alpha}{\partial t}|_{t=0}  =  \frac{\partial
\alpha}{\partial x}|_{t=0} = &0&  =
\frac{\partial^2\alpha}{\partial t \partial x}|_{P_0} =
\frac{\partial^2\alpha}{\partial x^2}|_{P_0}
\label{3}
\end{eqnarray}
Also,
\begin{equation}
\frac{2}{\alpha}\frac{ \partial^2 \alpha}{\partial t^2}  =  R
\label{4}
\end{equation}
where  $R$  is the curvature scalar.

The field equations in these coordinates, for a massless scalar field
read:
\begin{eqnarray}
\frac{1}{\sqrt{g}}\partial_\mu(
\sqrt{g}g^{\mu \nu}\partial_\nu)\phi&=&0 \nonumber \\
\frac{\partial^2 \phi}{\partial t^2 } +
\frac{\dot{\alpha}}{\alpha}\frac{\partial \phi }{\partial t}
+ \frac{\alpha'}{\alpha^3}\frac{\partial \phi }{\partial x }
- \frac{1}{\alpha^2}\frac{\partial^2 \phi }{\partial x^2 } &=&0.
\label{5}
\end{eqnarray}
Here,
\begin{equation}
\dot{\alpha}=\frac{\partial \alpha }{\partial t}
\ \ \ {\rm and}\ \ \ \alpha'=\frac{\partial \alpha }{\partial x}.
\end{equation}

The positive frequency modes $\phi$  of this field are obtained by
solving
these field equations with the two initial conditions
\begin{equation}
1)\ \ \ \ \ \ \ \ \ \      \phi_{p,\epsilon}(0,x) =
\frac{1}{\sqrt{4\pi p}}
\exp(- i p \epsilon x)\ \ \ \ \ \    p>0.
\label{6}
\end{equation}
Here we have
\begin{eqnarray}
\epsilon &=& +1\ \ \  {\rm corresponds\ \  to\ \  left\ \
travelling \ \
waves}\nonumber \\
\epsilon &=& -1\ \ \  {\rm corresponds\ \  to\ \  right\ \
travelling\ \
waves. }
\label{7}
\end{eqnarray}
and
\begin{equation}
2)\ \ \ \ \ \ \ \ \ \     i \frac{\partial
\phi_{p,\epsilon}}{\partial
t}|_{t=0}  =  p \phi_{p,\epsilon}|_{t=0}.
\label{8}
\end{equation}
A useful Ansatz to implement these initial conditions is:
\begin{equation}
\phi_{p,\epsilon}(t,x) = \frac{1}{\sqrt{4\pi p}} \exp(-i p
f_{\epsilon}(t,x))
\label{9}
\end{equation}
where  $f_\epsilon$ is real.
Equation (\ref{5}) then yields that
\begin{equation}
\frac{\partial  f_\epsilon}{\partial t}  =  \frac{\epsilon}{\alpha}
\frac{\partial  f_\epsilon}{\partial x}
\label{10}
\end{equation}
The initial conditions become
\begin{equation}
f_\epsilon(0,x) = \epsilon x
\label{11}
\end{equation}
and near  $t = 0$
\begin{equation}
 f_\epsilon(t,x) \approx t + \epsilon x
\label{12}
\end{equation}
The quantized field is now given by
\begin{equation}
\Psi(t,x) = \sum_{\epsilon=\pm 1}
\int_{0}^\infty d(\epsilon p)
\left(
\phi_{p,\epsilon}(t,x)a_{p,\epsilon} +
\phi_{p,\epsilon}^\ast(t,x)a_{p,\epsilon}^{\dagger}
\right)
\label{13}
\end{equation}
with the vacuum defined by
\begin{equation}
a_{p,\epsilon} |0>  = 0  .
\label{14}
\end{equation}
These modes have been normalized such that
\begin{eqnarray}
(\phi_{p,\epsilon} ,\phi_{q,\epsilon}) &=&  i\epsilon
\int_{-\infty}^{\infty}dx\sqrt{g}\left( \phi^{\ast}_{p,\epsilon}(t,x)
\stackrel{\leftrightarrow}{\partial_{t}}
\phi_{q,\epsilon}(t,x)
\right)\nonumber \\
&=& \frac{p+q}{4\pi\sqrt{pq}}\int_{-\infty}^{\infty}dx \epsilon
\alpha
\frac{\partial f_\epsilon }{\partial
t}\exp(i(p-q)f_\epsilon(t,x))\nonumber \\
&=& \frac{p+q}{4\pi\sqrt{pq}}\int_{-\infty}^{\infty}dx \frac{\partial
f_\epsilon }{\partial x}\exp(i(p-q)f_\epsilon(t,x))\nonumber \\
&=& \frac{p+q}{4\pi\sqrt{pq}} 2\pi \delta(p-q) \nonumber \\
&=&  \delta (p-q)
\label{15}
\end{eqnarray}

\section{The trace anomaly}
We begin with two ``observers"  with tangents to their world lines
given by
\begin{equation}
N^\mu(P_0) = (1,0)\ \ \      {\rm and}\ \ \     \bar{N}^\mu(P_0)
= (\cosh(\chi),\frac{\sinh(\chi)}{\bar{\alpha}})
\label{16}
\end{equation}
The corresponding frames are:
\begin{eqnarray}
e^{\mu \hat{0}} = (1,0)\ \ &\ &\ \ e^{\mu \hat{1}} = (0,
\frac{1}{\alpha})
\label{17} \\
\bar{e}^{\mu \hat{0}} =
(\cosh(\chi),\frac{\sinh(\chi)}{\bar{\alpha}})
\ \ &\ &\ \ \bar{e}^{\mu \hat{1}} =
(\sinh(\chi),\frac{\cosh(\chi)}{\bar{\alpha}})
\label{18}
\end{eqnarray}
Corresponding to this the metric has the two forms
\begin{equation}
ds^2 = dt^2 - \alpha^2(t,x) dx^2  = d\bar{t}^2 -
\bar{\alpha}^2(\bar{t},\bar{x})
d\bar{x}^2
\label{19}
\end{equation}

We can solve for the positive frequency modes in the barred as well
as in
the unbarred coordinates to obtain the corresponding quantized fields
$\bar{\Psi}(\bar{t},\bar{x})$ and $\Psi(t,x)$.  Their respective sets
of
annihilation and
creation operators are
$(\bar{a}_{p,\epsilon}  , \bar{a}^\dagger_{p,\epsilon})$  and
$(a_{p,\epsilon}
,a^\dagger_{p,\epsilon})$  .

	At $P_0$, the point with coordinates $(0,0)$ in both
coordinate systems the two fields coincide.
Corresponding to these two fields we
have their respective Fock space vacuums   $|\bar{0}>$  ,  $|0>$
defined by
\begin{equation}
\bar{a}_{p,\epsilon} |\bar{0}> = 0\ \ \  ,\ \ \  a_{p,\epsilon} |0> =
0
\label{20}
\end{equation}
Any bilinear expression in the field operators which, for physical
reasons, should have vanishing vacuum expectation value is defined by
normal ordering with respect to its own vacuum.  Thus since we expect
the
vacuum to be the state of zero energy and momentum density we require
that
\begin{equation}
<\bar{0}|:\bar{T}^{\hat{0}\hat{\mu}}:|\bar{0}> = 0
\label{20.5}
\end{equation}
and
\begin{equation}
<0|:T^{\hat{0}\hat{\mu}}:|0> = 0  ,
\label{21}
\end{equation}
where,
\begin{eqnarray}
T^{\hat{\alpha}\hat{\beta}} &=&
e^{\mu\hat{\alpha}}e^{\nu\hat{\beta}}T_{\mu\nu}
\nonumber \\
\bar{T}^{\hat{\alpha}\hat{\beta}} &=&
\bar{e}^{\mu\hat{\alpha}}\bar{e}^{\nu\hat{\beta}}\bar{T}_{\mu\nu}.
\end{eqnarray}

Furthermore, since the barred and unbarred frames  $\bar{e}^{\mu
\hat{\alpha}}$
,
$e^{\mu \hat{\alpha}}$ are related by a Lorentz transformation
\begin{equation}
\Lambda^{\hat{\beta}}_{ \ \hat{\alpha}} = \pmatrix{
\cosh(\chi) & \sinh(\chi) \cr
\sinh(\chi) & \cosh(\chi) \cr}
\label{22}
\end{equation}
we have that at $P_0$
\begin{equation}
:T^{\hat{\alpha} \hat{\beta}}:|_{P_0} =
\Lambda^{\hat{\alpha}}_{\ \hat{\gamma}}
\Lambda^{\hat{\beta}}_{ \ \hat{\delta}}
:\bar{T}^{\hat{\gamma}\hat{\delta}}:|_{P_0}
\label{23}
\end{equation}
so that in particular
\begin{equation}
:T^{\hat{0}\hat{0}}:|_{P_0} =
\cosh^2(\chi):\bar{T}^{\hat{0}\hat{0}}:|_{P_0} +
2\cosh(\chi)\sinh(\chi):\bar{T}^{\hat{0}\hat{1}}:|_{P_0} +
\sinh^2(\chi):\bar{T}^{\hat{1}\hat{1}}:|_{P_0}    .
\label{24}
\end{equation}
Taking the vacuum expectation value with respect to the barred vacuum
of
this equation, and using (\ref{20.5}) we have
\begin{equation}
<\bar{0}|:T^{\hat{0}\hat{0}}:|_{P_0}|\bar{0}> =  \sinh^2(\chi)
 <\bar{0}|:\bar{T}^{\hat{1}\hat{1}}:|_{P_0}|\bar{0}>
\label{25}
\end{equation}
Since  $<\bar{0}|:\bar{T}^{\hat{0}\hat{0}}:|\bar{0}> = 0$ we find
that the vacuum expectation value of the trace is:
\begin{equation}
<\bar{0}|\eta_{\hat{\alpha}\hat{\beta}} :
\bar{T}^{\hat{\alpha}\hat{\beta}}:|_{P_0}|\bar{0}> = -
<\bar{0}|:\bar{T}^{\hat{1}\hat{1}}:|_{P_0}|\bar{0}> = -
\frac{1}{\sinh^2(\chi)}
 <\bar{0}|:T^{\hat{0}\hat{0}}:|_{P_0}|\bar{0}>
\label{26}
\end{equation}
To evaluate this expression we have to take the term
$:T^{\hat{0}\hat{0}}:|_{P_0}$ which has
been normal ordered with respect to the vacuum  $|0>$ , rewrite it in
terms
of the operators
$(\bar{a}_{p,\epsilon},\bar{a}^\dagger_{p,\epsilon})$  and
commute the
terms so that the resulting expression  is normal ordered with
respect to the
vacuum  $|\bar{0}>$ .  To do this we write out the term
$:T^{\hat{0}\hat{0}}:|_{P_0}$
explicitly.  A simplification due to the use of equation (\ref{10})
occurs so
that only time derivatives of the field operators appear.  Also since
$\Psi(t,x)=\bar{\Psi}(\bar{t},\bar{x})$ we may
write
\begin{eqnarray}
:T^{\hat{0}\hat{0}}:|_{P_0} &=& :\frac{\partial \Psi}{\partial t}
\frac{\partial
\Psi}{\partial t}:|_{P_0}
= :\frac{\partial \bar{\Psi}}{\partial t} \frac{\partial
\Psi}{\partial
t}:|_{P_0} \nonumber \\
&=& \sum_{\epsilon=\pm 1} \int d(\epsilon p)
[\frac{\partial \bar{\Psi}}{\partial t}
\frac{\partial\phi_{p,\epsilon}}{\partial t} a_{p,\epsilon} +
\frac{\partial\phi^{\ast}_{p,\epsilon}}{\partial t}
a^\dagger_{p,\epsilon} \frac{\partial \bar{\Psi}}{\partial
t}]|_{P_0}.
\label{27}
\end{eqnarray}
To simplify the notation we drop the $|_{P_0}$ , but keep in mind
that these
equations only apply at  the point $P_0$.  Also we only evaluate this
expression for a fixed $\epsilon$.  Thus,
\begin{eqnarray}
:T_\epsilon^{\hat{0}\hat{0}}:  =  \int_{0}^{\infty} d(\epsilon p)
\int_{0}^{\infty} d(\epsilon q) \left[ \right. & &
(\frac{\partial \bar{\phi}_{q,\epsilon}}{\partial
t}\bar{a}_{q,\epsilon}+
\frac{\partial \bar{\phi}^{\ast}_{q,\epsilon}}{\partial
t}\bar{a}^{\dagger}_{q,\epsilon})a_{p,\epsilon}\frac{\partial
\phi_{p,\epsilon}}{\partial t}  \nonumber \\
&+& \left. \frac{\partial \phi^{\ast}_{p,\epsilon}}{\partial
t}a^{\dagger}_{p,\epsilon}
(\frac{\partial \bar{\phi}_{q,\epsilon}}{\partial
t}\bar{a}_{q,\epsilon}+
\frac{\partial \bar{\phi}^{\ast}_{q,\epsilon}}{\partial
t}\bar{a}^{\dagger}_{q,\epsilon})\right]
\label{28}
\end{eqnarray}
The operators $(a_{p,\epsilon} , a^\dagger_{p,\epsilon})$ are related
to the
barred
operators  $(\bar{a}_{k,\epsilon} ,\bar{a}^\dagger_{k,\epsilon})$  by
a
Bogolubov
transformation
\begin{equation}
a_{k,\epsilon} = \int d(\epsilon q)(\alpha_{k,q} \bar{a}_{q,\epsilon}
+
\beta^{\ast}_{k,q} \bar{a}^\dagger_{q,\epsilon})
\label{29}
\end{equation}
where
\begin{eqnarray}
\alpha_{k,q} &=& (\phi_{k,\epsilon} , \bar{\phi}_{q,\epsilon})
\nonumber \\
\beta_{k,q} &=& (\phi^{\ast}_{k,\epsilon} , \bar{\phi}_{q,\epsilon})
\label{30}
\end{eqnarray}

In our evaluation of the vacuum expectation value, the only term of
interest is the c-number term that results from the commutator
\begin{equation}
\bar{a}_{q,\epsilon'} \bar{a}^\dagger_{k,\epsilon} =
\bar{a}^\dagger_{k,\epsilon} \bar{a}_{q,\epsilon'} +
\delta_{\epsilon,\epsilon'} \delta(k-q)
\end{equation}
Thus, we get
\begin{equation}
<\bar{0}|:T_\epsilon^{\hat{0}\hat{0}}:|_{P_0}|\bar{0}> = \int
d(\epsilon p)
d(\epsilon q) [\frac{\partial \bar{\phi}_{q,\epsilon} }{\partial
t}
\frac{\partial \phi_{p,\epsilon}}{\partial t} \beta^\ast_{p,q}  +
c.c.  ]
\label{31}
\end{equation}
These terms are evaluated by replacing $\beta$ by its expression
(\ref{30}) and
interchanging the order of integration to first do the momentum
integrals.
In doing so the only regularization required is to define an integral
of
the form
\begin{equation}
\int_0^\infty dx x\exp(ixp)
\label{32}
\end{equation}
This is accomplished by replacing $p$ by $p + i \delta$ .  No further
regularizations are needed.  Further details of such a calculation
are in the appendix as well as the paper by Massacand and Schmid
\cite{1} and yield a Schwarz derivative.  The
final result is:
\begin{equation}
<\bar{0}|:T_\epsilon^{\hat{0}\hat{0}}:|_{P_0}|\bar{0}> =
\frac{1}{24\pi}
\frac{\frac{\partial^3\bar{f}_{\epsilon}}{\partial x^3}|_{P_0}}{\frac{\partial
\bar{f}_{\epsilon}}{\partial x}|_{P_0}}
\label{33}
\end{equation}
So we only have to evaluate these terms.  Now,
\begin{equation}
\frac{\partial \bar{f}_{\epsilon}}{\partial x} = \frac{\partial
\bar{f}_{\epsilon}}{\partial \bar{x}}
\frac{\partial \bar{x}}{\partial x} + \frac{\partial
\bar{f}_{\epsilon}}{\partial \bar{t}}
\frac{\partial \bar{t}}{\partial x}
\label{34}
\end{equation}
and as initial conditions at $P_0$ we have
\begin{equation}
\frac{\partial \bar{x}}{\partial x}|_{P_0} = \cosh(\chi)\ \ \ \     ,
\ \ \
\ \frac{\partial \bar{t}}{\partial x}|_{P_0}  = \sinh(\chi)
\label{35}
\end{equation}
Furthermore, we also have that
\begin{equation}
\bar{f}_\epsilon(0,\bar{x}) = \epsilon \bar{x}\ \ \ \   ,  \ \ \ \
\frac{\partial\bar{f}_{\epsilon}}{\partial \bar{t}} |_{P_0} = 1 \ \ \
 \  , \ \ \ \
\frac{\partial \bar{f}_{\epsilon}}{\partial \bar{x}}|_{P_0} = \epsilon
\bar{\alpha}\frac{\partial \bar{f}_{\epsilon}}{\partial \bar{t}}|_{P_0} =
\epsilon
\label{35.5}
\end{equation}
since  $\bar{\alpha}|_{P_0} = 1$  .
Also, as we stated earlier,
\begin{eqnarray}
\frac{\partial \bar{\alpha}}{\partial \bar{t}}|_{P_0} =
\frac{\partial
\bar{\alpha}}{\partial \bar{x}}|_{P_0} &=& 0
\label{36} \\
\frac{\partial^2\bar{\alpha}}{\partial \bar{t}\partial
\bar{x}}|_{P_0} =
\frac{\partial^2\bar{\alpha}}{\partial \bar{x}^2}|_{P_0} &=& 0
\label{37}
\end{eqnarray}
and
\begin{equation}
\frac{\partial^2\bar{\alpha}}{\partial \bar{t}^2}|_{P_0} =
\frac{R}{2}    .
\label{38}
\end{equation}
By repeatedly using the barred version of equation (\ref{10}) ,
namely
\begin{equation}
\frac{\partial \bar{f}_{\epsilon}}{\partial \bar{x}} = \epsilon \bar{\alpha}
\frac{\partial \bar{f}_{\epsilon}}{\partial \bar{t}}
\label{39}
\end{equation}
as well as
(\ref{34}), (\ref{35}) and (\ref{36}) we find:
\begin{equation}
\frac{\partial^2\bar{f}_{\epsilon}}{\partial \bar{t}^2}|_{P_0} =
\frac{\partial^2\bar{f}_{\epsilon}}{\partial \bar{t}\partial \bar{x}}|_{P_0} =
\frac{\partial^2\bar{f}_{\epsilon}}{\partial \bar{x}^2}|_{P_0} = 0
\label{40}
\end{equation}
as well as
\begin{equation}
\frac{\partial^3\bar{f}_{\epsilon}}{\partial \bar{x}^3}|_{P_0} = \epsilon
\frac{\partial^3\bar{f}_{\epsilon}}{\partial \bar{t}^3}|_{P_0} + \epsilon
\frac{\partial^2\bar{\alpha}}{\partial \bar{t}^2} \frac{\partial
\bar{f}_{\epsilon}}{\partial \bar{t}}|_{P_0} =0
\label{41}
\end{equation}
Thus we arrive at the result that
\begin{equation}
\frac{\partial^3\bar{f}_{\epsilon}}{\partial \bar{t}^3}|_{P_0} = -
\frac{\partial^2\bar{\alpha}}{\partial \bar{t}^2}|_{P_0} = -
\frac{R}{2}|_{P_0}
\label{42}
\end{equation}
This result now allows us to obtain that
\begin{eqnarray}
\frac{\partial^3\bar{f}_{\epsilon}}{\partial x^3}|_{P_0} &=&
[\frac{\partial^3\bar{t}}{\partial x^3} + \epsilon \bar{\alpha}
\frac{\partial^3\bar{x}}{\partial x^3} +
\epsilon \frac{\partial^2\bar{\alpha}}{\partial \bar{t}^2}
\frac{\partial
\bar{x}}{\partial x} (\frac{\partial \bar{t}}{\partial x})^2]
\frac{\partial
\bar{f}_{\epsilon}}{\partial \bar{t}}|_{P_0} +
(\frac{\partial \bar{t}}{\partial x} +\epsilon \frac{\partial
\bar{x}}{\partial
x}) \frac{\partial^3\bar{f}_{\epsilon}}{\partial \bar{t}^3} (\frac{\partial
\bar{t}}{\partial x})^2|_{P_0} \nonumber \\
&=& -\frac{R}{2} \sinh^3(\chi)  +
\frac{\partial^3\bar{t}}{\partial x^3} +
\epsilon \frac{\partial^3\bar{x}}{\partial x^3}
\label{43}
\end{eqnarray}
To evaluate the last two terms in this expression we use the fact
that
$(t,x)$ as well as $(\bar{t},\bar{x})$ satisfy the geodesic
equations, but have
different initial data on the spacelike geodesic that passes through
$P_0$.
These initial data are:
\begin{eqnarray}
\frac{d x}{ds}|_{P_0} = 1 \ \ \ \    &,& \ \ \ \
\frac{dt}{ds}|_{P_0} = 0
\label{44} \\
\frac{d\bar{x}}{ds}|_{P_0} = \cosh(\chi) \ \ \ \   &,& \ \ \ \
\frac{d\bar{t}}{ds}|_{P_0} = \sinh(\chi)
\label{45}
\end{eqnarray}
The geodesic equations read:
\begin{eqnarray}
\frac{d^2t}{ds^2} &=& - \alpha \frac{\partial  \alpha}{\partial t}
(\frac{dx}{ds})^2  , \nonumber \\
 \frac{d^2x}{ds^2} &=& - \frac{2}{\alpha}
\frac{\partial \alpha}{\partial t}\frac{dx}{ds} \frac{dt}{ds} -
\frac{1}{\alpha} \frac{\partial \alpha}{\partial x} (\frac{dx}{ds})^2
\label{46}
\end{eqnarray}
\begin{eqnarray}
\frac{d^2\bar{t}}{ds^2} &=& - \bar{\alpha} \frac{\partial
\bar{\alpha}
}{\partial \bar{t}} (\frac{d\bar{x}}{ds})^2  , \nonumber \\
 \frac{d^2\bar{x}}{ds^2} &=& - \frac{2}{\bar{\alpha}}
\frac{\partial \bar{\alpha}}{\partial \bar{t}}\frac{d\bar{x}}{ds}
\frac{d\bar{t}}{ds} - \frac{1}{\bar{\alpha}} \frac{\partial
\bar{\alpha}}{\partial \bar{x}} (\frac{d\bar{x}}{ds})^2
\label{47}
\end{eqnarray}
By differentiating  these equations as well as using  (\ref{35}) we
find that
\begin{equation}
\frac{\partial^2\bar{t}}{\partial x^2}|_{P_0} =
\frac{\partial^2\bar{x}}{\partial x^2}|_{P_0} = 0
\label{48}
\end{equation}
and
\begin{equation}
\frac{\partial^3\bar{t}}{\partial x^3}|_{P_0} = -
\frac{\partial^2\bar{\alpha}}{\partial \bar{t}^2}|_{P_0} \sinh(\chi)
\cosh^2(\chi) = - \frac{R}{2}
\sinh(\chi) \cosh^2(\chi)
\label{49}
\end{equation}
\begin{equation}
\frac{\partial^3\bar{x}}{\partial x^3}|_{P_0} = - 2
\frac{\partial^2\bar{\alpha}}{\partial \bar{t}^2}|_{P_0} \cosh(\chi)
\sinh^2(\chi)  = - R
\cosh(\chi) \sinh^2(\chi).
\label{50}
\end{equation}
Combining these results we obtain that
\begin{equation}
<\bar{0}|:T_{\epsilon}^{\hat{0}\hat{0}}:|_{P_0}|\bar{0}> =
-\frac{R}{48\pi}
\epsilon \exp(\epsilon \chi) \sinh(\chi)
\label{51}
\end{equation}
Adding the results for both values of $\epsilon$ we obtain
\begin{equation}
<\bar{0}|:T^{\hat{0}\hat{0}}:|_{P_0}|\bar{0}> =  -\frac{R}{24\pi}
\sinh^2(\chi)
\label{52}
\end{equation}
Inserting this into equation (\ref{25}) we finally obtain the vacuum
expectation
value of the trace of the stress-energy tensor, namely
$\frac{R}{24\pi}$ .

\section{Conclusion}
For the case of a conformally coupled massless scalar field in 1+1
dimensions it is much simpler to evaluate the  trace anomaly using a
particle picture than to avoid this.  The only regularization
required is
very simple, but it must be this very simple regularization that
suffices
to break the conformal symmetry and thus give a non-zero result for
the
vacuum expectation value of the trace.

\acknowledgments
AZC would like to thank the Theoretical Physics Institute of the
University of Innsbruck, as well as the Max-Planck-Inst. fuer Physik;
Werner Heisenberg Inst. for their hospitality.  He also acknowledges
support from the Natural Sciences and Engineering Research Council of
Canada (NSERC) as well as the Alexander von Humboldt Stiftung.

\appendix
\section*{}
We now evaluate explicitly the terms leading to the Schwartz derivative.
We have two terms, each of which is in turn a product of two factors.
Starting with expression (\ref{31}) we have,
\begin{equation}
<\bar{0}|:T_\epsilon^{\hat{0}\hat{0}}:|\bar{0}> = I = \int
d(\epsilon p)
\int d(\epsilon q) [\frac{\partial \bar{\phi}_{q,\epsilon} }{\partial
t}
\frac{\partial \phi_{p,\epsilon}}{\partial t} \beta^\ast_{p,q}  +
c.c.  ]
\label{A1}
\end{equation}
for $\beta$ we have
\begin{equation}
\beta^{\ast}_{p,q}=-i\epsilon \int_{-\infty}^{\infty} dy
\left( \phi^{\ast}_{p,\epsilon}(y) \partial_t
\bar{\phi}^{\ast}_{q,\epsilon}(\bar{y})
- \partial_t \phi^{\ast}_{p,\epsilon}(y)
\bar{\phi}^{\ast}_{q,\epsilon}(\bar{y})
\right)
\label{A2}
\end{equation}
where we have dropped the $\alpha = 1$, because we perform the integral on
the $t=0$ surface. This leaves us with
\begin{eqnarray}
I&=& -i\epsilon \partial_t\bar{f}(\bar{x})\int_{-\infty}^{\infty} dy
\int d(\epsilon p) \partial_t \phi_{p,\epsilon}(x) \phi_{p,\epsilon}^{\ast}(y)
\int d(\epsilon q) \partial_{\bar{t}} \bar{\phi}_{q,\epsilon}(\bar{x})
\partial_{t} \bar{\phi}_{q,\epsilon}^{\ast}(\bar{y}) \nonumber \\
&+&i\epsilon\partial_t\bar{f}(\bar{x}) \int_{-\infty}^{\infty} dy
\int d(\epsilon p) \partial_t \phi_{p,\epsilon}(x) \partial_t
\phi_{p,\epsilon}^{\ast}(y)
\int d(\epsilon q) \partial_{\bar{t}} \bar{\phi}_{q,\epsilon}(\bar{x})
\bar{\phi}_{q,\epsilon}^{\ast}(\bar{y}).
\label{A3}
\end{eqnarray}
We must now perform the following integrals,
\begin{eqnarray}
\int d(\epsilon p) \partial_t \phi_{p,\epsilon}(x)
\phi_{p,\epsilon}^{\ast}(y)&=&
\frac{-i}{4\pi}\int_0^{\infty} dp e^{-i p(x-y-i\delta)}
\nonumber \\
&=&\frac{-1}{4\pi}\frac{1}{(x-y-i \delta)}
\nonumber \\
\int d(\epsilon q) \partial_{\bar{t}} \bar{\phi}_{q,\epsilon}(\bar{x})
\partial_{t} \bar{\phi}_{q,\epsilon}^{\ast}(\bar{y})&=&
\frac{\epsilon}{4\pi}\int_0^{\infty} dq q \partial_t
\bar{f}_{\epsilon}(\bar{y})
e^{-i\epsilon q( \bar{f}_{\epsilon}(\bar{x})-
\bar{f}_{\epsilon}(\bar{y}) -i\epsilon\delta)}
\nonumber \\
&=&\frac{-\epsilon}{4\pi}
\partial_t \bar{f}_{\epsilon}(\bar{y}) \frac{1}{(\bar{f}_{\epsilon}(\bar{x})-
\bar{f}_{\epsilon}(\bar{y})-i\epsilon\delta)^2}
\nonumber \\
\int d(\epsilon p) \partial_t \phi_{p,\epsilon}(x) \partial_t
\phi_{p,\epsilon}^{\ast}(y)&=&
\frac{\epsilon}{4\pi}\int_0^{\infty} dp p e^{-ip(x-y-i\delta)}
\nonumber \\
&=& -\frac{\epsilon}{4\pi}\frac{1}{(x-y-i\delta)^2}
\nonumber \\
\int d(\epsilon q) \partial_{\bar{t}} \bar{\phi}_{q,\epsilon}(\bar{x})
\bar{\phi}_{q,\epsilon}^{\ast}(\bar{y})&=&
\frac{-i}{4\pi}\int_0^{\infty} dq e^{-i\epsilon q( \bar{f}_{\epsilon}(\bar{x})-
 \bar{f}_{\epsilon}(\bar{y}) -i\epsilon\delta)}
\nonumber  \\
&=&-\frac{\epsilon}{4\pi}\frac{1}{(\bar{f}_{\epsilon}(\bar{x})-
\bar{f}_{\epsilon}(\bar{y})-i\epsilon\delta)}.
\label{A4}
\end{eqnarray}
Using these results we are left with
\begin{eqnarray}
I=(\cosh \chi + \epsilon \sinh \chi)\frac{-i}{16\pi^2}
\int_{-\infty}^{\infty} dy \frac{(x-y)^2}{
(\bar{f}_{\epsilon}(\bar{x})-\bar{f}_{\epsilon}(\bar{y}))^2} \partial_t
\bar{f}_{\epsilon}(\bar{y}) \frac{1}{
(x-y-i\epsilon\delta)^3} +c.c.
\nonumber \\
+(\cosh \chi + \epsilon \sinh \chi)\frac{i}{16\pi^2}
\int_{-\infty}^{\infty} dy
\frac{\epsilon(x-y)}{(\bar{f}_{\epsilon}(\bar{x})-
\bar{f}_{\epsilon}(\bar{y}))}
 \frac{1}{(x-y-i\epsilon\delta)^3} + c.c. .
\label{A5}
\end{eqnarray}
Using the identity
\begin{equation}
i\pi \delta''(y-x) = \frac{1}{(x-y-i\delta)^3} -
\frac{1}{(x-y+i\delta)^3}
\label{A6}
\end{equation}
we now have
\begin{equation}
I=(\cosh \chi + \epsilon \sinh \chi)\frac{1}{16\pi}\int_{-\infty}^{\infty}
\delta''(y-x) \left(
\frac{-\epsilon(x-y)}{\bar{f}_{\epsilon}(\bar{x})-
\bar{f}_{\epsilon}(\bar{y})} + \partial_t \bar{f}_{\epsilon}(\bar{y})
\frac{(x-y)^2}{(\bar{f}_{\epsilon}(\bar{x})-\bar{f}_{\epsilon}(\bar{y}))^2}
\right).
\label{A7}
\end{equation}
To simplify this expression we use the relations
\begin{eqnarray}
\lim_{y \rightarrow x}\partial^2_y \frac{x-y}{
\bar{f}_{\epsilon}(\bar{x})-\bar{f}_{\epsilon}(\bar{y})}&=&
\frac{(\bar{f}_{\epsilon}'')^2}{2( \bar{f}_{\epsilon}')^3} -
\frac{\bar{f}_{\epsilon}'''}{3( \bar{f}_{\epsilon}')^2} \\
\lim_{y \rightarrow x}\partial^2_y \left(g(y)\frac{(x-y)^2}{
(\bar{f}_{\epsilon}(\bar{x})-\bar{f}_{\epsilon}(\bar{y}))^2}\right)&=&
-\frac{2g'\bar{f}_{\epsilon}''}{ (\bar{f}_{\epsilon}')^3} + \frac{3g(
\bar{f}_{\epsilon}'')^2}{2(\bar{f}_{\epsilon}')^4} +\frac{g''}{
(\bar{f}_{\epsilon}')^2} -
\frac{2g\bar{f}_{\epsilon}''' }{3(\bar{f}_{\epsilon}')^3}.
\label{A8}
\end{eqnarray}
where $g=\partial_t \bar{f}_{\epsilon}(\bar{y})$ and the dashes represent
partial differentiation with respect to the unbarred coordinates not the barred
coordinates.
It is easy to show that $\bar{f}_{\epsilon}''=0$. This leaves us with
\begin{equation}
I=(\cosh \chi + \epsilon \sinh \chi)\frac{1}{16\pi}\left(
\frac{\epsilon \bar{f}_{\epsilon}'''}{3(\bar{f}_{\epsilon}')^2}
+\frac{g''}{(\bar{f}_{\epsilon}')^2} - \frac{2 g
\bar{f}_{\epsilon}'''}{3(\bar{f}_{\epsilon}')^3}
\right)
\label{A9}
\end{equation}
to get the result we're after we factor out a factor of
$\frac{\bar{f}_{\epsilon}'''}{\bar{f}_{\epsilon}'}$ and
use the relationship \\ $g''|_{P_0}=\epsilon \bar{f}_{\epsilon}'''|_{P_0}$,
this leaves us with
\begin{equation}
I= (\cosh \chi + \epsilon \sinh \chi) \frac{1}{16\pi}
\frac{\bar{f}_{\epsilon}'''}{\bar{f}_{\epsilon}'}\left( \frac{\epsilon}{3f'} +
\frac{\epsilon}{\bar{f}_{\epsilon}'}-
\frac{2 g}{3(\bar{f}_{\epsilon}')^2}
\right)
\label{A10}
\end{equation}
which when the last factor is evaluated at $P_0$ yields,
\begin{eqnarray}
I&=& (\cosh \chi + \epsilon \sinh \chi) \frac{1}{24\pi} \frac{
\bar{f}_{\epsilon}'''}{\bar{f}_{\epsilon}'}
(\cosh \chi - \epsilon \sinh \chi) \nonumber \\
&=& \frac{1}{24\pi} \frac{\bar{f}_{\epsilon}'''}{\bar{f}_{\epsilon}'} |_{P_0}
\label{A11}
\end{eqnarray}

\end{document}